# Fixed point of self-similar Lennard-Jones potentials in the glass transition

**Jialin Wu**

College of Material Science and Engineering, Donghua University,

Shanghai, 200051, China, Email: jlwu@dhu.edu.cn

The existence of potential energy landscape in the glass transition has been theoretically proved using the recursive equation of reinforce-restraint of self-similar 8 orders of Lennard-Jones (L-J) potential fluctuations. The stability condition for fluctuation reinforce-restraint is just the Lindemann ratio that is exactly deduced as 0.1047 in this paper. The origin and transfer of interface excitation comes of the balance between self-similar L-J potential fluctuation and geometric phase potential fluctuation, which also gives rise to a new attractive potential of $-17/16\varepsilon_i$, lower than the potential well energy $-\varepsilon_i$ of $i$-th order of L-J potential, in the self-similar mean field of mean fields of different sizes. The delocalization energy of two-body is exactly the transfer energy of excited interface, and the delocalization path is along 8 orders of geodesics in topological analyses.

In the previous paper [1], a theoretical framework has been proposed that the glass transition (GT) is only determined by the intrinsic 8 orders of instant 2-$D$ mosaic geometric structures, without any presupposition and relevant parameter. According to the theory of thermally excited ripplon [2] and the results of [3], the interface excitation between two reference particles in GT has been illustrated. An interface excitation is a vector with 8 orders of relaxation times, 8 orders of additional restoring torques, quantized interface excited energy (QIEE) of $1/8\varepsilon_i$ ($\varepsilon_i$ is potential well energy of $i$-th order of two body interaction) and extra volume. Each order of harmonic restoring torque gives rise to an additional position-asymmetry of Lindemann displacement [4, 5]; thus, in order to eliminate the additional position-asymmetry, the accompanied 8 orders of transient 2-$D$ clusters with the 4 neighboring interface relaxations of the reference particle have been formed. Here, a transient 2-$D$ cluster is strictly defined by an instant interface excited symmetric loop on projection plane. In order to step by step eliminate the anharmonic interface tensions, the 8 orders of dynamical 3-$D$ hard-spheres should be adopted. It is interesting that the physical picture provides a unified mechanism to interpret hard-sphere [6], compacting cluster [7], non-ergodic [8], free volume [9], cage [10], percolation [11], instantaneous normal mode [12], geometrical frustration [7] and jamming behaviors [13]. It is proposed [1] that the clusters in mosaic structures turn out to be the special hard-spheres with the quantized interface interaction potential of $1/8\varepsilon_i$ and the appearance of 8 orders of hard-spheres respectively correspond to the 8 orders of local cluster growth phase transitions (LCGPT) in GT.

In this paper, the independent testification for fixed point of self-similar Lennard-Jones (L-J) potentials will directly educe that there are only 8 orders of self-similar hard-spheres in GT and explain the origins of QIEE and the potential energy landscape [3, 14]. In addition, one of the most prominent open questions in GT concerns the so-called tunneling in two-level [3], which will be unambiguously explained by using the theory of fixed point of L-J potentials in this paper.

The classical L-J potential that accords with the scaling theory in critical phase transition is used to discuss the GT.

$$U_{1,2}(\sigma_i, q) = q^x f(\sigma_i / q) = 4\varepsilon_i [(\frac{\sigma_i}{q})^{12} - (\frac{\sigma_i}{q})^6] \tag{1}$$

In Fig.1, due to the left-right asymmetry of potential curve (1) with regard to the balance position $q_{1,0}$, two clusters with $\sigma_1$ have unequal amplitudes on the two sides in $q$-axial. The two $q$ values on the two sides of $U$ are respectively denoted as $q_{1,L}$ and $q_{1,R}$.

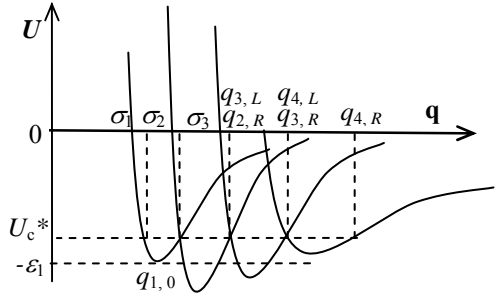

**Figure 1.** At the equilibrant point of attraction and repulsion, $U_c$, ($U_c \in (0, \varepsilon_i)$), there are no end of $U_c$ points and self-similar L-J potential curves satisfying $U_c = U(q_{i,R}) = U(q_{i+1,L})$. They are all unstable except for curves via the fixed point, $U_c^*$, of L-J potentials in GT.

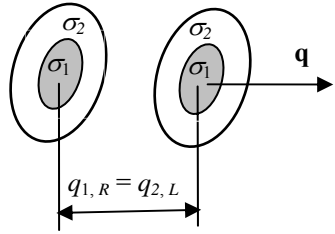

**Figure 2.** Self-similar clusters of two different sizes satisfy $U_c = U(q_{1,R}) = U(q_{2,L})$ along $q$-axis, the direction of cluster growing. The two $\sigma_1$ in $\sigma_2$ clusters are in 'balance state' of attraction and repulsion during the 2nd order of LCGPT, when the QIEE transfer from inside $\sigma_2$ to the surface of $\sigma_2$.

In GT, the fractal vibration appears as is shown in Fig.2, in which the two $\sigma_1$ clusters (black sphere) are apart $q_{1,R}$ and in attractive state. On the other hand, these two clusters also respectively belong to two larger $\sigma_2$ clusters at the instant time of the 2nd order of LCGPT. As long as these two $\sigma_2$ clusters are also apart $q_{1,R}$ but on the right side of the 2nd order L-J potential in Fig.1, i.e. $q_{1,R} = q_{2,L}$, these two $\sigma_1$ in $\sigma_2$ clusters are also in repulsive state. The potential that the attraction and repulsion happens to be in equilibrium is denoted as $U_c$.

By recursively applying this procedure, $M$ orders of clusters can be obtained from small to large in size along the q-axis seeing Fig.1, i.e. there are infinite self-similar L-J potential curves satisfying $U_c = U(q_{i,R}) = U(q_{i+1,L})$, where $U_c$ is in $(0, \varepsilon_i)$. These equilibrium points of attraction and repulsion at $q_{iR}$, the sharp-angled points in Fig.1, are unstable in general. The critical phase transition always occurs on an unstable critical point. Among all the possible values of $U_c$, there is, however, a point which is not only an unstable critical point in LCGPT but also a stable bifurcation point. We denote this point as $U_c^*$ which corresponds to the *intrinsic stability condition* in solid →liquid GT. Let $q_{i,R}^*$ and $q_{i+1,L}^*$ satisfy: $U_c^* = U(q_{i,R}^*) = U(q_{i+1,L}^*)$. Based on the scaling theory, the $(\sigma/q)$ in eq (1) has to be constant. Hereby a characteristic parameter is introduced

$$\chi = (\sigma/q)^6 \quad (2)$$

It will be interpreted later on that $\chi$ represents the geometric phase potential formed by the non-integrable geometric phase factor between local (two-body interaction) and global ($2\pi$ cycle interaction of reference cluster surrounded by neighboring clusters), and is called *reduced geometric phase potential* (RGPP) in this paper, which is independent of temperature.
Equation (1) is rewritten as

$$U_{1,2}(\chi)/\varepsilon_i = f(\chi) = -4\chi(1-\chi) \quad (3)$$

As $q$ increases, the tow-body interaction potential *does not necessarily increase* in GT, see Fig.1. Due to the left-right asymmetry of potential curve (1), $\chi$ in eq (3) must have two solutions, which correspond to two different constants in eq (2), denoted respectively as $\chi_{+1}$ and $\chi_{-1}$ (subscript represents slope, see eq (9) and Fig. 4).

The density fluctuation stability in mode-coupling theory can be here regarded as the stability of compacting *i*-th order clusters with *lower density* and re-building (*i*+1)-th order with *higher density* in the LCGPT, seeing [1]. Suppose there is a disturbance of the RGPP arising from the displacive (or distance-) fluctuation in compacting *i*-th order clusters, denoted as $\Delta\chi_i$, there is a corresponding disturbance in re-building (*i*+1)-th order clusters, denoted as $\Delta\chi_{i+1}$, $\Delta\chi_{i+1}$ then vice versa adds to its interaction potential $f(\chi_i)$ to excite disturbed *i*-th order cluster as an additional excited fluctuation potential. The recursion equation is given by

$$f(\chi_i) + \Delta\chi_{i+1} = f(\chi_i + \Delta\chi_i) \quad i=1, 2, \ldots \quad (4)$$

Eq (4) comes of : $f(\chi_i) + \Delta f(\chi_i) = f(\chi_i + \Delta\chi_i)$ and the recursive rule $\Delta\chi_{i+1} = \Delta f(\chi_i)$. Notice that the type of the recursion equation is in nature different from that in current literature [15]. The recursively defined variable is the potential fluctuation $\Delta\chi$ in RGPP$\chi$, whose recurrent motion of 'fist reinforce – after restraint' let $\chi$ finally recursive in $\chi_{+1}$ or $\chi_{-1}$. Here the key idea is as that of the picture of additional replica symmetry breaking in GT described in previous paper [1]: on projection plane perpendicular to *q*-axial (this means the interface excitations on a $2\pi$ closed loop belong to *parallel transport* [16] surround *q*-axial in topology analysis), only by fluctuations of excited interfaces may a $2\pi$ interface excited loop of (*i*+1)-th order 2-D cluster ($\sigma_{i+1}$) appear at the instant time $t_{i+1}$ in reference $a_0$ local field to eliminate the additional position-asymmetry of any *i*-th order 2-D cluster ($\sigma_i$) during ($t_i, t_{i+1}$), however, because of the jamming effect [1], this procedure is in fact retarded until the forming condition of 8th order loop is satisfied, hereafter, the 8 orders of $2\pi$ closed loops will one after another appear and let the mass centers of the 8 orders of 2-D clusters one by one return to the vibrant balance position of reference $a_0$ particle on projection plane. Because of a $2\pi$ instant closed loop on projection plane corresponding a singularity on *q*-axial, the singularity is exactly the non-integrable geometric phase factor [16] (being of geometric phase potential), here comes the restricted relationship of geometric phase factors between local and global ($2\pi$ cyclic in parallel transport) in each order loop, as also, the *i*-th order loop and (*i*+1)-th order loop, or in other words, the relationship between mean field $U(\sigma_i)$ and mean field $U(\sigma_{i+1})$ in equation (1). The recursion equation eq (4) is a mean field equation of mean fields in different size, which represents the restricted relationship. That is, eq (4) is a recursion equation containing RGPP fluctuation stability in LCGPT. The recursive procedure of eq (4) is that a disturbance $\Delta\chi$ on the bifurcation point, $\chi_{+1}$, via first fluctuation reinforce and after fluctuation restraint, finally corresponds to the recurrent point $\chi_{-1}$, and a disturbance $\Delta\chi$ on the bifurcation point $\chi_{-1}$ to the recurrent point $\chi_{+1}$, which will be explained in Fig.4.

Notice again that in random system, all 2-D local excited fields are in random orientations, thus, using statistical 3-D clusters instead of 2-D clusters in Eq (1) does not influence our discussions, whereas, the evolution characteristics of hard-spheres with various sizes along a same direction and the non-integrable geometric phase potential of $2\pi$ closed loop surrounding $q$-axial should be still reserved.

Using the linear approximation:

$$f(\chi_i + \Delta\chi_i) = f(\chi_i) + \frac{\partial f(\chi_i)}{\partial \chi_i} \Delta\chi_i \tag{5}$$

If there exists the fixed point of $\chi_i$, denoted as $\chi^*$, eqs (3), (4) and (5) may be rewritten as

$$\Delta\chi_{i+1} = \left. \frac{\partial f(\chi)}{\partial \chi} \right|_{\chi=\chi^*} \cdot \Delta\chi_i \tag{6}$$

$$f(\chi^*) = -4\chi^*(1-\chi^*) \tag{7}$$

Equation (7) is the fixed point equation of self-similar L-J potential functions. From (6):

$$\frac{\Delta\chi_{i+1}}{\Delta\chi_i} = \left. \frac{\partial f(\chi)}{\partial \chi} \right|_{\chi=\chi^*} \tag{8}$$

For stable fixed point, the absolute value of $\Delta\chi_{i+1}$ must be less than that of $\Delta\chi_i$. Thus, the stability condition of fixed point can be similarly derived as literature [15]

$$s \equiv \left| \frac{\Delta\chi_{i+1}}{\Delta\chi_i} \right| = \left| \frac{\partial f(\chi)}{\partial \chi} \right|_{\chi=\chi^*} \leq 1 \tag{9}$$

From (7) and (9), it can be identified that $3/8 \leq \chi \leq 5/8$. In LCGPT, all RGPP fluctuate between the two bifurcation points of $\chi$, respectively the minimum value 3/8, corresponding to compacting into $i$-th order clusters on sharp-angled points in Fig.1, and the maximum value 5/8, corresponding to forming $(i+1)$-th order clusters. According to the recursive rule, for the solutions of eq (9), $\chi_{+1} = 3/8$ and $\chi_{-1} = 5/8$ (the solution for slope $-1$ in eq (9) recursive in 5/8, see Fig.4) are adopted. Thus, there are only two types of stable RGPPs: $\chi_{+1} = 3/8$ stands for the faster RGPP, which is contributed by the attraction of the $i$-th order clusters in $(i+1)$-th order, and $\chi_{-1} = 5/8$ stands for the slower RGPP, which is contributed by the repulsion of the $(i+1)$-th order clusters containing $i$-th order.

By substituting $\chi_{+1}$ or $\chi_{-1}$ into (7), the sole fixed point of L-J potentials can be obtained

$$U_c^* = -15/16\,\varepsilon_i \tag{10}$$

In addition, $\chi_{+1} + \chi_{-1} = 3/8 + 5/8 = 1$, which means that (because of $f(\chi_{+1} + \chi_{-1}) = 0$), on each sharp-angled point in Fig.3, the sum of the two fast-slow geometric phase potentials exactly equals to the cluster evolution energy $\varepsilon_i$ of one exterior degree of freedom, which is consistent with the result of [1]. By substituting $\chi_{+1}$, $\chi_{-1}$ into (3), it can be concluded that there are only two stable RGPPs during the LCGPT, though there can be $M$ solutions for degenerate state:

$$\chi_{+1} = (\sigma_i / q_{i,R})^6 \quad \text{and} \quad \chi_{-1} = (\sigma_{i+1} / q_{i+1,L})^6, \; i = 1,2\ldots M \tag{11}$$

satisfying $\chi_{+1} = 3/8$, $\chi_{-1} = 5/8$.

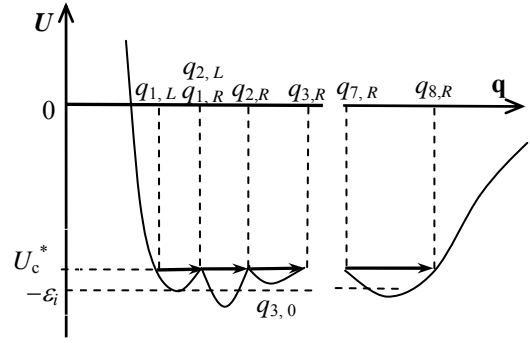

**Figure 3.** At the fixed point in GT, $U_c^*$, there are 8 sharp-angled points, $q_{i,R}$, forming the apart (re-coupling) paths of two $\sigma_1$ clusters along 8 orders of geodesics, from $q_{1,L} \rightarrow q_{1,R} \rightarrow q_{2,R} \rightarrow q_{3,R} \ldots \rightarrow q_{8,R}$, geodesic is the shortest line between $q_{i,R}$ and $q_{i+1,R}$ on the cylindrical potential surface formed by $U_c^*$ surround axis $U = -\varepsilon_i$ a cycle.

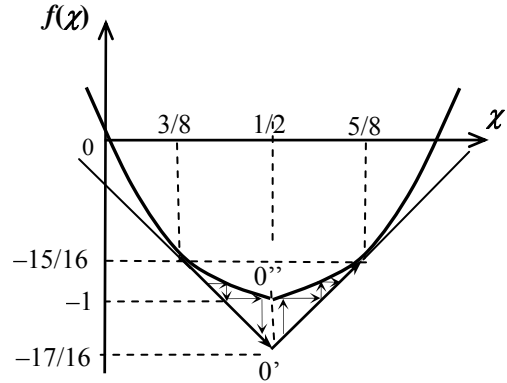

**Figure 4.** During GT, the path of compacting cluster starts from a disturbance $\Delta\chi$ on bifurcation point (3/8, −15/16), via fluctuation reinforce losing one QIEE, to the point 0' along slope −1, then via fluctuation restraint gaining one QIEE, to the bifurcation point (5/8, −15/16) along slope 1, in which one QIEE of $1/8\varepsilon_i$ transfers from the surface on $i$-th order cluster to that on $(i+1)$-th order. The attractive potential on point 0' comes from the self-similar fluctuation equation $\Delta f / f \equiv \Delta\chi / \chi$.

The stability condition of recursion equation eq (4) can be also deduced from the simple self-similar condition of eq (12) in the new coordinate system (transformation: $\chi' = \chi - 1/2$, $f' = f + 17/16$) of 0' (1/2, -17/16) as coordinate origin in Fig. 4. Eq (12) and the difference of 17/16 −15/16 = 1/2 − 3/8 = 5/8 − 1/2 = 1/8 (the QIEE) in Fig.4 clearly give voice to that the fixed point $U_c^*$ in Fig.1 and the origin of interface excitation and the transfer of excited interface all result from the balance effect between the two-body interaction potential fluctuation and the cluster geometric phase potential fluctuation in GT. This is also the bifurcation stability condition in eq (4). Fixed point $U_c^*$ is easy to be proved since in figure 1, none of the potential curves, except for those who have passed, correspond with the stability condition of eq (12).

$$\frac{\partial f'}{\partial \chi'} \equiv \frac{f'}{\chi'} \tag{12}$$

Three important results thus are found. (i) There are two 'attraction-repulsion' balance points in GT, one is the point 0''(1/2, −1) in Fig.4, corresponding to the points $(q_{i,0}, \varepsilon_0)$ in Fig.3, that of the solid state syntonic vibrations of clusters, and the other is the point (1/2, −17/16) in Fig.4, corresponding to the balance both the RGPP fluctuation and two-body interaction potential fluctuation at point $\chi = 1/2$. (ii) The apart path of two $\sigma_1$ clusters from $q_{i,R}$ to $q_{i+1,R}$ in Fig.3 surrounds (do not pass!, refracts the parallel transport surround $q$-axial in topology analysis) the vibrate balance point $q_{i,0}$ in Fig.3, i.e. the point 0" in Fig.4, and the sum of the two orthogonal geometric phase potentials on sharp-angled point $q_{i,R} = q_{i+1,L}$ equals to the potential well energy, which clearly indicate the arrow from $q_{i,R}$ to $q_{i+1,R}$ in Fig.3 is exactly the non-integrable geometric phase factor in multiparticle system [17] or the geodesics [18], the minimum energy path, in topological analysis. (iii) The two-body interactions of two $\sigma_1$-clusters have 8 orders interacting (relaxation) times and the 8 orders interaction potential of $1/8\varepsilon_i$ in their potential energy landscape.

The potential energy $-17/16\varepsilon_i$ is a newly arisen additional attractive potential in the mean field (Eq (12)) of mean fields of different sizes (Fig.1), which further resolves the puzzle why colloidal hard-sphere system has stronger attractive potential. The singularity of the mean field of mean fields of different sizes can be seen in $q_{i,R} = q_{i+1,L}$ (verified by eqs (13)- (16)) that the right displacive amplitude of $q_{i,R}$ in $i$-th order mean field balances the left displacive amplitude of $q_{i+1,L}$ in ($i+1$)-th order, instead of the left-right asymmetric vibrant in potential curve (1), thus this displacive amplitude energy of particle-cluster is the additional in the glass transition.

'Self-similar fluctuation' in eq (12) itself is also a kind of 'ordering' relative to random fluctuation. It is interesting that the positions of two attraction potentials of $-15/16\varepsilon_i$ and $-17/16\varepsilon_i$ are in reflection symmetry relative to that of the potential well energy in Fig.4. This further explains that the origin of quantized interface excitation: interface excitation not only balances both the RGPP fluctuation and two-body interaction potential fluctuation, but also balances both the additional self-similar attractive potential and the additional displacive amplitude energy in the self-similar mean fields of different sizes in the glass transition, instead of so-called "two-level". The abnormal additional displacive amplitude energy in self-similar system may be the causation of abnormal heat capacity and abnormal heat expansion in the glass transition.

From $\chi_{+1} = 3/8$, $\chi_{-1} = 5/8$, $\sigma_i$ and $q_{i,R}$ in Fig. 3 can respectively be expressed as

$$q_{i,R} = q_{i+1,L} \tag{13}$$

$$\sigma_i = \sigma_1 (5/3)^{(i-1)/6}, \quad i = 1, 2, 3\ldots \tag{14}$$

$$q_{i,R} = \sigma_1 (5/3)^{(i-1)/6} (8/3)^{1/6} \tag{15}$$

$$\Delta q_{i+1} = q_{i+1,R} - q_{i+1,L} = q_{i+1,R} - q_{i,R} = (8/3)^{1/6} \left[ (5/3)^{1/6} - 1 \right] \sigma_i \approx 0.1047 \sigma_i \tag{16}$$

In Eq. (15), when $i = 8$, $q_{8,R} \approx 2.137\sigma_1 > 2\sigma_1$, ($q_{7,R} < 2\sigma_1$). Therefore, only at the 8th order L-J potential field is the 8th order 'distance-increment vacancy' with excluded volume of $\sigma_1$ able to be put in between two $\sigma_1$ clusters with same excluded volume. In the meantime, when $i = 8$, the system also satisfies the classical condition ($v_c = 3b$, $b$ here is $\sigma_1$) of critical phase transition of the Van der Waals long range interaction equation ($(P + a/v^2)(v - b) = kT$). However, the critical energy to form the 8th order cluster is the localized energy, $E_c = kT_g^* = 20/3\varepsilon_1$ [1], less than the critical phase transition energy, $kT_c = 8\varepsilon_1$, in Van der Waals equation. This is clearly also the contribution of the long-range correlation of short-range interaction of excited interfaces in LCGPT. The result of $M = 8$ further confirms that there are only 8 orders of intrinsic mosaic geometric structures [1] and 8 orders of LCGPT in GT.

It can be seen that the 8 orders of distance-increment vacancies in GT respectively offer one outer degree of freedom for 8 orders of cluster motions in inverse cascade, which come from the 8 orders of geometric phase potentials formed by 8 orders of $2\pi$ interface excited closed loops, independent of temperature and frequency. This characteristic may be used to understand the origin of Boson peak [7].

There are two degenerate states for RGPP on each sharp-angled point in Fig.3. The apart paths of two $\sigma_l$ clusters show in Fig. 4: inside $\sigma_{i+1}$, $\sigma_i$ cluster is of 5 inner degrees of freedom [1] and can migrate in the range of $\sigma_{i+1}$; thus, the slow process re-coupling or delocalization of two $\sigma_l$-clusters should be adopted accompanied with $\sigma_8$ cluster repetitious rearrangements. The delocalization energy is only $1/8\varepsilon_i$, the quantized transfer energy of interface excitation. Therefore, the 8 orders of apart (delocalization) paths of two-body and the local structure repetitious rearrangements and the attractive potential $-17/16\varepsilon_i$ reveal the mechanisms of the so-called "tunneling" [19, 20] in GT.

Eq (16) also shows that the stable ratio of $d_L = \Delta q_{i+1}/\sigma_i = 0.1047\ldots$, a universal value for real linear polymer, which is exactly the Lindemann ratio. Lindemann criterion has been applied to study solid-versus-liquid behavior [21]: distance-fluctuations cannot increase without destroying the lattice structure [21]. Here it has been found that the Lindemann ratio contained in eq (16) is equivalent to the stability condition of recursion equation eq (4), which also indicates LCGPT in GT is only determined by the intrinsic geometric structure, without any presupposition and relevant parameter. The numerical value of $d_L$ directly deduced from this paper accords with the approximation value of 0.10 in [22] and 0.1- 0.15 in [21]. This also validates the theory of fixed point for self-similar L-J potentials and the eq (13).

In summary, one of the most striking is that the existence of fixed point for self-similar L-J potentials can be actually proved without other assumption and complicated mathematical analysis. However, the stable fluctuation recursion equation in GT reveals the relationship between 'reinforce - restraint' of potential-fluctuation and Lindemann ratio. This theory provides a unified mechanism to interpret the Lindemann ratio, the origin of quantized interface excitation, and the 8 orders of hard-spheres in GT, the hard-sphere long-range attractive potential and the tunneling. The accurate Lindemann ratio has been directly obtained as 0.1047. A universal behavior in GT is found that two orthogonal degenerate states of fast-slow geometric phase factors are accompanied with the appearance of each order of LCGPT. The fixed point also clarifies that the so-called tunneling turns out to be the generating and transferring of all quantized interface excitations should pass through the additional attractive potential center of $-17/16\varepsilon_i$, which is lower than potential well energy and comes from the balance effect of reinforce-restraint of self-similar potential fluctuation. Finally, the delocalization path or the re-coupling path of particle-pairs is along 8 orders of geodesics and the delocalization energy only as $1/8\varepsilon_i$ of L-J potential well energies, which is exactly the quantized interface excited energy.
.

**Acknowledgements:** President Yuan Tseh Lee and Professor Sheng Hsien Lin of the Academia Sinica, Taiwan supported the manuscript preparation. Valuable discussion with Professor Yun Huang of Beijing University, Professor Da-Cheng Wu of Sichuan University, Professor Bo-Ren Liang and Professor Cheng-Xun Wu of Donghua University are also acknowledged.